\documentstyle[11pt,aaspp4]{article}
\newcommand{\Ind}{\hbox{Ind}}
\newcommand{\bz}{{\bar z}}
\newcommand{\bff}{{\bf f}}

\begin{document}


\title{General Theorem about Gravitational Lensing}
\author{TAKESHI FUKUYAMA\altaffilmark{1}}
\affil{Department of Physics, Ritsumeikan University, Kusatsu,
	Shiga, 525 Japan}

\and

\author{TAKASHI OKAMURA\altaffilmark{2}}
\affil{Department of Physics, Tokyo Institute of Technology, 
	Oh-okayama, Meguro-ku, Tokyo, 152 Japan}

\altaffiltext{1}{fukuyama@bkc.ritsumei.ac.jp}
\altaffiltext{2}{okamura@th.phys.titech.ac.jp}
\begin{abstract}
We study the general theorem about gravitational lensing 
which states the relationship between 
the numbers of images with different parities. 
Our formulation allows an extension to the nontransparent 
and singular model.
\end{abstract}

\section{Introduction}
There are the general theorems about the gravitational lensing%
(\cite{MacKenzie} ; \cite{Schneider})
which mainly state the relationships between the numbers of the images 
with different parities;
\begin{equation}
	n_I-n_{II}+n_{III}=1 \, ,
\label{index}
\end{equation}
where $n_I$ is the number of images of type I.
We use $n_{II}$ and $n_{III}$ analogously.

The types of images are defined by the Jacobian matrix of 
the Fermat's potential $\phi$,
\begin{equation}
	A_{ij} \equiv {\partial^2 \phi \over 
	       \partial x_i \partial x_j } \qquad (i,j=1,2) \, .
\label{Jacobian}
\end{equation}
\begin{description}
\item[Type I] : both eigenvalues of the Jacobian matrix are positive.
\item[Type II] : one eigenvalue is positive and another negative.
\item[Type III] : both are negative.
\end{description}
Therefore type I and type III images have positive parities and
type II images have negative ones.

The existing general theorems are general for transparent and 
non-singular lens models.
Typical examples are the elliptical lens models 
with softened power law behavior%
(\cite{Binney} ; Schneider, et al.\ 1992).
However, they do not exhaust all the lens models. 
They are rather phenomenological models, whereas 
there is a widely valid model.
It is multipole expansion model(MPE model) 
which is obtained by expanding deflection potential 
with respect to impact parameter%
(\cite{Yoshida} ; \cite{Kakigi} ; \cite{Fukuyama}).
The latter is less popular than the former in lensing analysis
but is very powerful
(Fukuyama, et al.\ 1997).
The MPE model is nontransparent and singular, 
and we see how the general theorems are extended 
to incorporate MPE model and other models.  
In this extension we propose the generalized index theorem 
valid also to the nontransparent and singular lens model and
the formulation in complex coordinates is found 
to be very convenient.
\section{Formulation} \label{formulation}
Before proceeding to the concrete physical situation, 
let us begin with the mathematical preparations. 
Index theorem for critical points in two dimensional flat or
closed curved surfaces is well known for mathematicians
(\cite{Eguchi}).
They are summarized as follows: 
We consider the vector field on the surface $S$.  
Denoting the critical points (at which the vector field vanishes) 
by $P_i$ on $S$, we get the following theorem.
\begin{equation}
	\sum_i ( \mbox{Index of the vector at } P_i)=\chi (S) \, ,
\label{Euler}
\end{equation}
where $\chi (S)$ is the Euler characteristic of $S$.  
For flat space $\chi$ is one and the result becomes equation (\ref{index}).  
%
%
This theorem does not hold in the case that the vector field has 
singular points. 
So we reformulate the index theorem so as to incorporate pole terms. 

Let us consider a complex function $f(z)$ 
with a complex variable $z$ that is not necessarily holomorphic. 
The function $f$ plays the role of the vector field 
mentioned above. 
The index of $f$ is defined by
\begin{equation}
	\Ind (f) \equiv {1\over 2\pi i} \oint_{C_z} 
	          d\, \ln \left( f \left(z \right) \right) \, .
\label{def}
\end{equation}
This measures the variation of the phase of $f$ 
along the closed curve $C_z$ which goes anti-clockwise.
When $C_z$ does not enclose any zero points or poles,
the index vanishes.
If $C_z$ encloses some zero points or poles, $\{P_i \}$,
then the index is given by the summation of the contributions from
each point $P_i$,
\begin{eqnarray}
	\Ind (f) &=& \sum_i \Ind (f~;P_i) \, ,
\label{indsum} \\
	\Ind (f~;P_i) &=& {1 \over 2 \pi i} \oint_{C^{(i)}_z}
	          d\, \ln \left( f \left(z \right) \right) \, ,
\label{indeach}
\end{eqnarray}
where $C^{(i)}_z$ is the circle enclosing the point $P_i$.
In the context of the gravitational lenses, $f$ is replaced by 
$\phi_\bz$($\equiv \partial\phi/\partial\bz$; 
$\bz$ is complex conjugate of $z$) and $\phi_\bz=0$ is just lens equation 
(see eqs.[\ref{finGL}] and [\ref{Fermat}]).
Therefore ordinary images are zero points of order one.
If lens object has finite total mass and extended regular mass density, 
then we have the pole only at infinity.
Whereas in the case of MPE models and 
models with singular mass density such as point masses, 
other poles are produced.

{}From the form of equation (\ref{indsum}) we should pay attention 
to each contribution.
Let us consider the map $\omega =f(z)$ 
in the neighborhood of $P_i$.
Then the $\Ind (f~;P_i)$ is rewritten as
\begin{equation}
	\Ind (f~;P_i)={1 \over 2\pi i} \oint_{C_{\omega}} 
	d\, \ln \omega \, ,
\end{equation}
where $C_{\omega} =\omega (C^{(i)}_z )$.  
The value of the right hand side is
the (algebraic) number of times that we wind around $C_{\omega}$ 
when we go around $C^{(i)}_z$ in the anti-clockwise direction.

Firstly we consider zero point case and proceed to the case of poles. 
Hereafter we set the origin of complex $z$-plane 
at the zero point $P_i$ and we use bold-face letters for the value
of any function $g$, for example, ${\bf g}=g(P_i)$.
%
\subsection{The case of zero point of order one} \label{orderone}
%
In the neighborhood of a zero point, $\omega$ can be written as
\begin{equation}
	\omega = z \bff_z + \bz \bff_\bz +O(|z|^2) \, ,
\end{equation}
where lower indices $z$ and $\bz$ mean partial derivatives 
with respect to them, for example, 
$$
	f_{z^k \bz^l} \equiv 
	{ {\partial}^{k+l} \over \partial z^k \partial \bz^l }\, f 
	\, .
$$

We define $\mu$ by 
$$
	\mu \equiv {\bff_\bz \over \bff_z} \, .
$$  
Then the Jacobian $D$ of the mapping from $z$ to $\omega$ is 
in this case given by
\begin{equation}
	D=\bigl| \bff_z  \bigr|^2 - \bigl| \bff_\bz \bigr|^2 
	= \bigl| \bff_z \bigr|^2 \bigl( 1-|\mu|^2 \bigr) \, .
\label{jacobian}
\end{equation}
That is, $\omega=f(z)$ reserve (reverse) 
the direction of rotation if $|\mu|< 1$ ( $|\mu|> 1$ ).
\par
Winding number of $C_{\omega}$ is one for the point of order one and
\begin{equation}
	\Ind (f~;P_i)=  \left\{ 
	\begin{array}{ll}
		1  &   \mbox{for  } |\mu| < 1 \\
        	-1 &   \mbox{for  } |\mu| > 1 \, .
\end{array}
\right.
\end{equation}
%
\subsection{The case of zero point of higher orders}
%
About a zero point, $\omega$ is expanded as a polynomial of $z$ and $\bz$.  
However the winding number is determined by the lowest power term as 
is easily checked.  

For the simplicity, we consider the case that dominant part
in the neighborhood of the zero point consists of the only term as
\begin{equation}
	\omega ={\bff_{z^k \bz^l} \over k!~ l!} z^k \bz^l
	+(\mbox{subdominant terms})
\label{power}
\end{equation}
We call this type of zero point as $(k,l)$-type critical point
for later use.  
Although equation (\ref{power}) restricts the form of $f$,
as we see later, for the case of point masses and MPE models,
it is enough to examine equation (\ref{power}).

Then the index of $f$ is given by 
\begin{equation}
	\Ind (f~;P_i)=k-l \, .
\end{equation}
Therefore, from the previous subsection \ref{orderone} ,
we can regard zero points with $|\mu| < 1$ as $(1,0)$-type and
ones with $|\mu| > 1$ as $(0,1)$-type.
%
\subsection{The case of pole terms}
%
As was mentioned, besides the zero points, pole terms of $f$ also 
contribute to $\Ind (f)$.  
In the neighborhood of pole of $f$ 
we define $g$ by $g \equiv 1/f$, then
\begin{equation}
	\Ind (f~;P_i)=- \Ind (g~;P_i)
\end{equation}
comes from the definition of equation (\ref{def}).  
If $f$ consists of several pole terms, $\Ind (f)$ is determined 
by the highest pole term.

Thus, similarly to the case of zero points,
we can regard the pole whose dominant term has 
the form of equation (\ref{power}) 
as $(k,l)$-type critical point with negative integers $k$ and $l$.
\section{Application to gravitational lensing}
We have completed mathematical preparations and proceed to 
application to gravitational lensing.  
In this case $f$ is defined through the Fermat's potential 
$\phi$ by
\begin{equation}
	f=\phi_\bz \, ,
\label{finGL}
\end{equation}
where
\begin{equation}
	\phi ={1\over 2} |z-z_s|^2 -\psi (z) \, .
\label{Fermat}
\end{equation}
$z(z_s)$ are image (source) positions in the complex notations and 
$\psi$ is the deflection potential defined by 
\begin{equation}
	\psi (z) \equiv {1 \over \pi}\int d^2 z' \kappa (z') \ln |z-z'| \, ,
\end{equation}
where the integral measure $d^2 z$ means $d\Re(z) d\Im(z)$ and
$\kappa$ is normalized surface mass density of lens object.
We assume 
\begin{equation}
	\lim_{|z| \to \infty} \left| {\psi_\bz \over z} \right|=0 \, ,
\end{equation}
and this condition may hold for the models with finite total mass.

For the large impact parameter, $\phi_\bz$ becomes
\begin{equation}
	\phi_\bz = {1\over 2}(z-z_s)+o \left( |z| \right) \, .
\end{equation}
That is, infinite point becomes pole.
Remarking that the anti-clockwise loop $C_z$ ($|z| \gg 1$) 
is viewed as clockwise one from the infinite point, 
we obtain the contribution of the infinite point to the index as
\begin{equation}
	\Ind (\phi_\bz)=1 \, .
\label{fermat}
\end{equation}
Diminishing the radius of the circle $C_z$, 
it is entangled with the zero points and poles.  
Then equation (\ref{fermat}) must be equal to 
the sum of these contributions.

As we have mentioned, the zero point of order one 
in the neighborhood of the origin is classified into  
$(1,0)$-type  for $|\mu|<1$ and 
$(0,1)$-type for $|\mu|>1$ by the contribution to the index.
Thus, we can say that the contribution comes from the term 
$$
	\omega \sim {C_{k,l} \over k!~ l!} z^k \bz^l \, ,
$$
where $C_{k,l}$ is a constant and positive (negative) $k$ and $l$ correspond 
to the zero points (poles). 
We denote the number of $(k,l)$-type critical points as 
$n^{(k,l)}$.  
Then from the arguments of the section \ref{formulation}, we get
\begin{equation}
	1=\Ind (\phi_\bz)=\sum_{k,l} (k-l) \, n^{(k,l)} \, .
\end{equation} 
\begin{list}{$\bullet$}{\setlength{\leftmargin}{15pt}}
\item For the extended lens models which have no pole we obtain
\begin{equation}
	1=n^{(1,0)}-n^{(0,1)}=n_+-n_-=n_I+n_{III}-n_{II} \, ,
\end{equation}
where $n_+$($n_-$) is the number of the images 
with the positive(negative) parity.
\item  For the point masses,
we have a pole of $(0,-1)$-type for every point mass and therefore
\begin{equation}
1=n_I +n_{III}-n_{II}+ \left( \mbox{\# of masses} \right) \, .
\end{equation}
\item  For a MPE model up to $2^k$-pole,
Fermat's potential is given by%
(Kakigi, et al.\ 1995)
\begin{equation}
	\phi ={1\over 2}|z-z_s|^2-{1\over 2}\ln (z \bz)
	+ \sum_{n=1}^k \left( 
	{Q_n \over z^n} + {{\bar Q_n} \over {\bar z}^n} 
	\right)
\end{equation}
and the vector field becomes
\begin{equation}
	\phi_\bz = {1\over 2}(z-z_s)-{1 \over 2 \bz}
	  -{1 \over 2} \sum_{n=1}^k \left( n 
	{ {\bar Q_n} \over \bz^{n+1} } \right) \, .
\end{equation}
So there give rise a pole of $(0,-k-1)$-type and 
\begin{equation}
	1=n^{(1,0)}-n^{(0,1)}+(k+1)~n^{(0,-k-1)}
	 =n_I+n_{III}-n_{II}+k+1 \, .
\label{new}
\end{equation}
\end{list}

\begin{figure}
\plotone{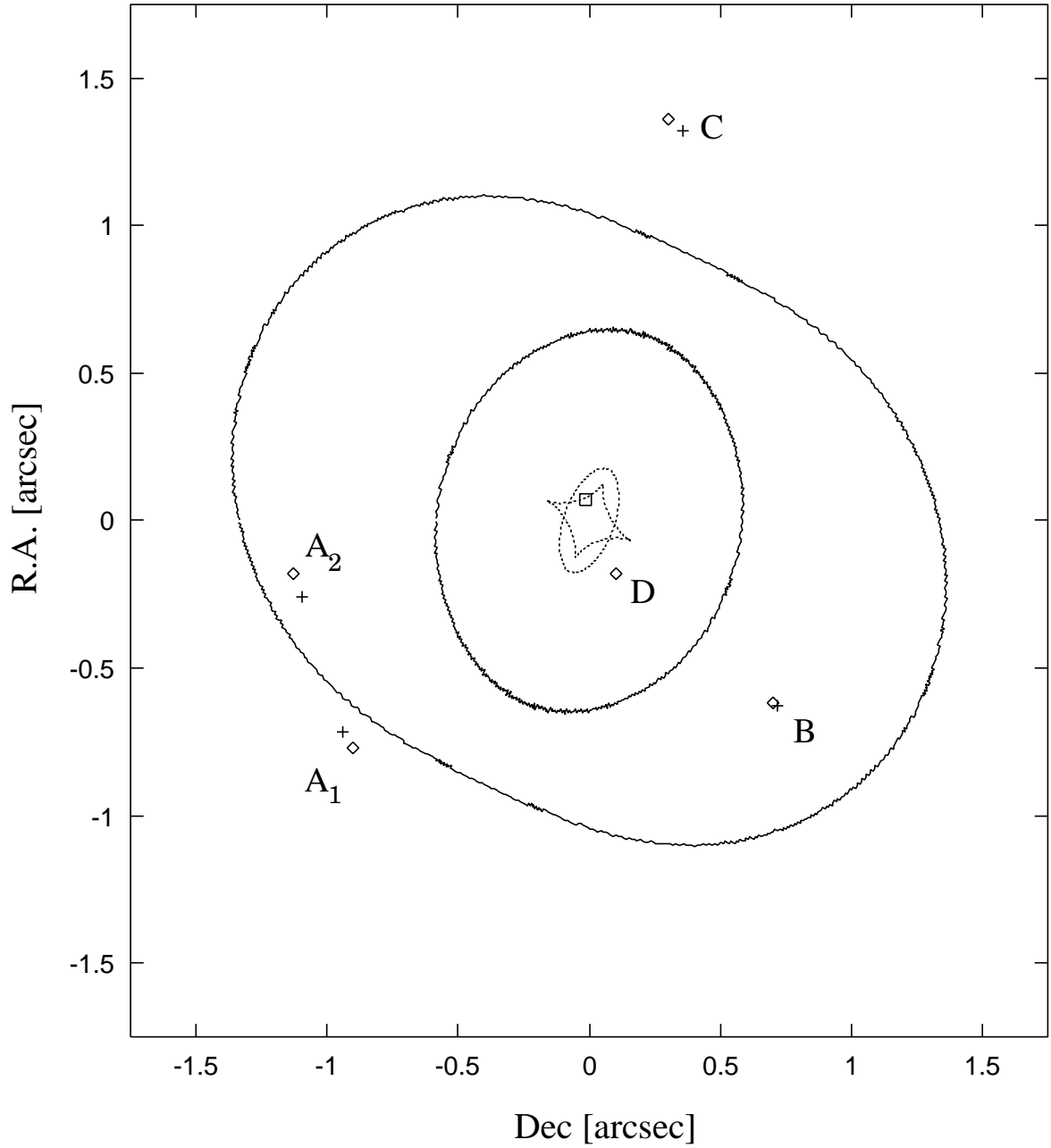}
\caption{The observed image positions (crosses) and numerically 
fitted ones(diamonds) by a model with softened power law surface mass density.
The square shows source position. The solid and dotted lines are
critical line and caustics, respectively.
\label{fig1}}
\end{figure}

\begin{figure}
\plotone{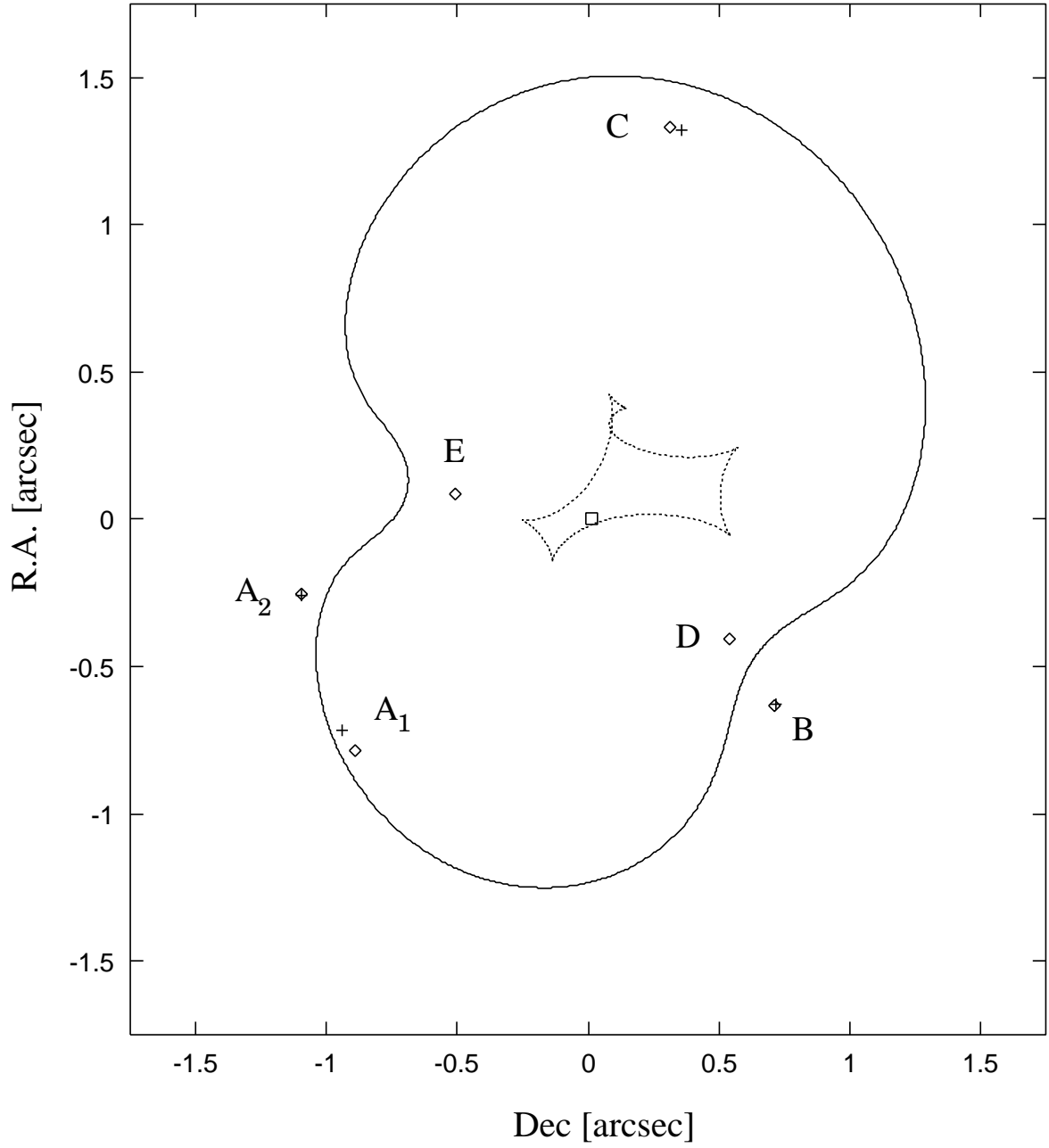}
\caption{ibid. by a MPE model. 
\label{fig2}}
\end{figure}

Lastly, we give the concrete example of the theorem 
in the multiple images of PG1115+080%
(\cite{Young} ; \cite{Christian} ; \cite{Kristian} ; \cite{Schechter}).
\placefigure{fig1}
\placefigure{fig2}
Fig.~\ref{fig1} shows critical and caustics lines of the Fermat's potential
and image positions numerically fitted 
by a model with softened power law surface density
(\cite{Asano}).
Critical lines are those on which Jacobian determinant equation 
(\ref{Jacobian}) vanishes.
So the parities of images are reversed across a critical line.
Images outside the outermost critical line have positive parities.
Hence critical lines are useful for determination of parities of images.

From positions of the critical lines, we can know that 
$A_1$ and $C$ belong to $n_I$,
$A_2$ and $B$ do to $n_{II}$ and ${D}$ to $n_{III}$. 
So the calculated numbers of images satisfy the original theorem
equation (\ref{index}).

On the other hand Fig.~\ref{fig2} shows fitting of the same lens system by
a MPE model including up to quadrupole
(Fukuyama, et al.\ 1997).
$A_2$ and $B$ belong to $n_I$.
The other $A_1$, $C$, $D$ and $E$ do to $n_{II}$. 
Thus, these satisfy equation (\ref{new}) 
for the case of quadrupole, $k=2$.
%
\acknowledgments
The authors would like to thank Professor E.L. Wright for comments.
 
\newpage
\end{document}